\documentstyle[preprint,aps]{revtex}

\def\epem{e^+ e^-}
\def\mpmm{\mu^+ \mu^-}
\def\ppbar{p\bar{p}}
\def\muToeee{\mu^- \rightarrow e^- e^+ e^-}
\def\muToegam{\mu^- \rightarrow e^- \gamma}
\def\mueToee{e^{\mp} \mu^{\pm} \rightarrow e^+ e^-}
\def\mueToff{e^{\mp} \mu^{\pm} \rightarrow f_1 f_2 }
\def\eeTomue{e^+ e^- \rightarrow \mu^{\pm} e^{\mp}}
\def\GeV{{\rm GeV}}
\def\TeV{{\rm TeV}}

\input{epsf}

\begin{document} 

\vspace*{-1.1cm}
\begin{flushright}          
\hfill{YUMS 97-21}\\
\hfill{SNUTP 97-095}\\
\hfill{DESY 97-142}\\
%\hfill{hep-ph/9707483}\\
\hfill{\today}
\end{flushright}

\vspace{0.5cm}

\begin{center}
{\Large \bf High Energy FCNC search through $e\mu$ Colliders} %\\
%{\Large \bf $\mu \rightarrow e \gamma$, $\mu \rightarrow e e e$, 
%$e e \rightarrow \mu e$ {\it vs.} $e \mu \rightarrow f_1f_2$}

\vspace{0.8cm}
 
{\large 
S. Y. Choi\footnote{sychoi@theory.yonsei.ac.kr},
~~C. S. Kim\footnote{kim@cskim.yonsei.ac.kr,
   ~cskim@kekvax.kek.jp, ~http://phya.yonsei.ac.kr/\~{}cskim/}, 
~~Y. J. Kwon\footnote{kwon@phya.yonsei.ac.kr}
~~and~~Sam-Hyeon Lee
}\\

{\it Department of Physics, Yonsei University,
Seoul 120-749, Korea}\\

\vspace{1.4cm}
\end{center}

\begin{center}
{\bf Abstract} 
\end{center}

\noindent
We study the potential impacts of a new type of particle collider --
an $e\mu$ collider -- on the search for new physics beyond the
Standard Model.  As our first attempt for exploring its physics
potential, we demonstrate that the the $e\mu$ collision experiment can
be highly efficient in searching for lepton-number-violating Flavor
Changing Neutral Current phenomena.

\vfill

\pagebreak 

Although it is possible to explain the observed CP violation\cite{cronin}
within the framework of the Standard Model (SM), it is generally
believed that the amount of CP asymmetry predicted by the SM is
insufficient to explain the observed non-zero baryon asymmetry in the
universe, which inevitably requires a much larger extent of CP
asymmetry \cite{Huet}.  Consequently, it is expected that there must
be new physics beyond the SM in the high energy regime, such as SUSY,
GUT {\it etc.}  One of the key signatures for such new physics is the
lepton-number-violating Flavor Changing Neutral Current (FCNC)
phenomenon.
There have been numerous theoretical studies on
lepton-number-violating FCNC by using new models, {\it e.g.} the
generalized two Higgs doublet model \cite{g2hdm}, as well as by
considering various collider and decay processes \cite{col}.
Moreover, there have been experimental studies from Los Alamos, CERN
{\it etc.}, on low-energy reactions \cite{Bolton,Bellgardt},
$\muToegam$, $\muToeee $.  All these muon decay experiments, however,
are limited within the low energy regime by the muon's small rest
mass.  Therefore, even if any FCNC effects due to an unknown massive
neutral particle exist, the effects would be severely suppressed due
to its high virtuality.

In this paper, we show that the $e\mu $ collision, which is very
similar to the above mentioned $\mu $ decay reactions, can be a
powerful alternative to explore such FCNC phenomena.  
The $e\mu $ collision, in connection with the problem of
muonium-antimuonium transitions \cite{mam}, $\mu^+ e^- \leftrightarrow
\mu^- e^+$, through doubly charged Higgs, $\Delta^{++}$, dilepton
gauge boson, $X^{++}$, or flavor changing neutral scalar bosons, $H$
and $A$, has been first illustrated by Hou \cite{hou}.
By using a simple model-independent calculation, we demonstrate that
the $e\mu$ collision experiment can be much more efficient than the
present low-energy rare $\mu $ decay experiments and $\epem$ collision
experiments, such as $\muToegam$, $\mu\rightarrow eee$ and
$e^+e^-\rightarrow e^\pm\mu^\mp$, in searching for
lepton-number-violating FCNC phenomena.  In the SM, the probability of
$\muToegam$ and $\mu\rightarrow eee$ are absolutely zero, while their
experimental upper limits are $4.9\times 10^{-11}$ \cite{Bolton} and
$1.0\times 10^{-12}$ \cite{Bellgardt}, respectively.  The $\epem $ and
$\ppbar $ collisions which have been most powerful tools in the
high-energy particle physics experiments, or even a $\mpmm $
collision\cite{Barger} which is being studied for future experiments,
are dominated by the SM interaction via such well-known particles as
$\gamma $ (photon) or $Z^0$.  On the contrary, the $s$-channel $e \mu$
interaction can never be mediated by $\gamma$ or $Z^0$, and hence is
very sensitive to the effects of new and unknown neutral particles.

The tree level FCNC phenomena can be detected
straightforwardly in $e\mu$ collisions through
\begin{displaymath}
\mueToff,~~f_{1,2}={\rm
lepton,~quark,~gauge~boson,~supersymmetric~particle~{\it etc.},}
\end{displaymath}
{\it i.e.}
through any two-body (or more-body) final state  
except for a few channels allowed in the SM.
We explain several important advantages of the 
$e \mu$ collision experiment over the existing methods such as  
$\mu \rightarrow e\gamma$, $\mu\rightarrow eee$, and $\eeTomue$:

(i) $e \mu \rightarrow f_1 f_2$ vs $\mu\rightarrow e\gamma$:
~~The advantage of $e \mu$ collision is obvious in this comparison.  
While the latter reaction can allow us 
to detect FCNC caused only through photon mediation, 
the $e \mu$ collision experiment enables us to detect FCNC
not only through $\gamma$ or $Z$ mediation but also through the
exchange of a new neutral particle, {\it e.g.} a supersymmetric
Higgs, neutralino, scalar or vector GUT gauge boson, and so on.

(ii) $e \mu \rightarrow f_1 f_2$ vs $\mu\rightarrow eee$:
~~The $e \mu$ collision experiment, where $e$ and $\mu$ can be
accelerated up to very high energies, has great advantages for
new physics at high-energy scales which can not be directly
probed by the latter low-energy process $\mu \rightarrow eee$.
In this light, the $e \mu$ collision
experiment becomes much more powerful for a heavier particle mass
scale.  Needless to say, by lowering beam energy, we can also
investigate the process equivalent to $\mu \rightarrow eee$, thus
providing a cross-check for the FCNC search results obtained from
low-energy $\mu $ decay experiments.

(iii) $e \mu \rightarrow f_1 f_2$ vs 
$e^+e^-\rightarrow \mu^\pm e^\mp$:
~~The two processes have an identical physics origin, if $f_1 f_2 =
e^+ e^-$.  The $e \mu$ collision experiment has, however, a large
number of channels such as $e \mu \rightarrow ee, \mu\mu,
\tau\tau,\tau e, \tau\mu,$ $u\bar{u}, d\bar{d}, s\bar{s}, c\bar{c},
b\bar{b}, t\bar{t}, u\bar{c},$ $u\bar{t}, c\bar{t}, d\bar{s}, d\bar{b},
s\bar{b}, c\bar{u}$, ..., $gg$, $W^+W^-$, ..., and so on, while the
$ee$ collision experiment has only three relevant processes,
$ee\rightarrow \mu e$, $ee\rightarrow\tau e$ and $ee\rightarrow
\tau\mu$.  Moreover, once we consider the color factor $N_C=3$, and
the production of gauge bosons, supersymmetric particles or possible
new scalar and vector boson pair productions, then the number of
available channels becomes even larger.  Another advantage of $e\mu
\rightarrow f_1 f_2$ reaction over $e^+e^-\rightarrow \mu^\pm e^\mp$
is that muons can be easily accelerated to very high energies without
much synchrotron radiation loss because of their large mass compared
to that of an electron, thus making $e \mu$ collider a much better
option than the conventional $\epem$ collider to reach ultra high
energy regime for FCNC search.  

To make a simple comparison of $e^\pm\mu^\mp\rightarrow f_1 f_2$ mode
to the other cases, let us assume that the FCNC is mediated by an
unknown neutral boson $X$ of mass $M_X$ and width $\Gamma_X$.  When
the electron mass is neglected, the decay width of $\mu^- \rightarrow
e^- e^+ e^- $ is given by
\begin{equation}
\label{Gamsme3}
\Gamma (\mu^- \rightarrow e^- e^+ e^-) = {m_\mu^5 \over {2048 \pi^3}}
\left| {g_{e\mu}^{^S} g_{ee}^{^S} \over M_X^2 } \right|^2,
\end{equation}
for a scalar boson $X$, while the decay width becomes
\begin{equation}
\label{Gamvme3}
\Gamma (\mu^- \rightarrow e^- e^+ e^-) = {m_\mu^5 \over {384 \pi^3}}
\left| {g_{e\mu}^{^V} g_{ee}^{^V} \over M_X^2 } \right|^2,
\end{equation}
for a vector boson $X$. Here $g_{e\mu}^{^S}$ ($g_{ee}^{^S}$) 
and $g_{e\mu}^{^V}$ ($g_{ee}^{^V}$) are  
the appropriate flavor-changing 
(flavor-conserving) coupling strength of leptons with scalar $X$ 
and vector $X$, respectively.  In both cases, the total decay width 
decreases by $1/M_X^4$, assuming that the coupling strengths are 
independent of $M_X$. 

On the other hand, the cross-section of 
$e^\pm\mu^\mp\rightarrow e^+e^-$ takes the following form
(we simply chose the case $f_1 f_2 = e^+ e^-$): 
\begin{itemize}
\item{For a scalar $X$,
\begin{equation}
\sigma_{_S} (\mueToee) = 
    {|g_{e\mu}^{^S} g_{ee}^{^S} |^2 \over 16 \pi s} 
\left[ \left| \Pi (s) \right|^2 - 
{\rm Re} [ \Pi (s) ]L(\xi)
+ 2\left( L(\xi)-{1 \over \xi+2 }\right)
\right], 
\end{equation}
where $\xi=2M^2_X/s$, and the two functions 
$\Pi(s)$ and $L(\xi)$ are  defined as
\begin{eqnarray*} 
\Pi (s) =\frac{s}{s-M^2_X+i\sqrt{s}\Gamma_X},\qquad
L(\xi)  =1-{\xi \over 2}\log\left({\xi +2 \over \xi }\right).
\end{eqnarray*} 
The peak cross-section of $e^\pm\mu^\mp\rightarrow e^+e^-$ 
at $s=M^2_X$ is then given by
\begin{equation}
\sigma_{_S} (\mueToee) |_{s=M_X^2} = {|g_{e\mu}^{^S} g_{ee}^{^S} |^2 
\over 16 \pi \Gamma_X^2 } \left[ 1+ 2\left(\Gamma_X\over M_X\right)^2 
\left( {3 \over 4} - \log 2 \right) \right].
\end{equation}
}
\item{For a vector $X$,
\begin{equation}
\sigma_{_V} (\mueToee)  =  
                 {|g_{e\mu}^{^V} g_{ee}^{^V} |^2 \over 32 \pi s} 
\left[ {8 \over 3} \left| \Pi (s) \right|^2 - {1\over 2} {\rm Re} 
[ \Pi (s) ] A(\xi) + B(\xi) 
\right], 
\end{equation}
where the functions $A(\xi)$ and $B(\xi)$ are given by
\begin{eqnarray*}
&& A(\xi)=2(3+\xi)-(2+\xi)^2 \log\left({\xi+2 \over\xi }\right),\\ 
&& B(\xi)=2\left[ 4 + {8\over \xi} - {4\over {\xi+2}} -2(\xi+2)
          \log\left({\xi +2 \over \xi }\right) \right].
\end{eqnarray*}
The corresponding peak cross-section at $s=M^2_X$ is then 
\begin{equation}
\sigma_{_V} (\mueToee) |_{s=M_X^2} = {|g_{e\mu}^{^V} g_{ee}^{^V} |^2 
\over \pi \Gamma_X^2} \left[ {1 \over 12} + 
{1\over 16}\left( \Gamma_X\over M_X\right)^2 
\left( 7 - 8 \log 2 \right) \right] .
\end{equation}
}
\end{itemize}
Note that in a sharp contrast to $\Gamma (\muToeee)$,
which decreases as $1/M_X^4$, the peak cross-section $\sigma
(\mueToee) |_{s=M_X^2}$ is simply proportional to $1/\Gamma^2_X$
if $\Gamma_X \ll M_X$. As $\Gamma_X/M_X$ is typically $10^{-2}$-$10^{-3}$
for a weakly decaying $X$, it is obvious that the $e\mu $ collision 
experiment can be much more efficient in the search for FCNC phenomena 
mediated by a heavy neutral particle than the low-energy rare 
decay process, $\mu\rightarrow eee$.
 
\begin{figure}
\leavevmode
\epsfysize=5.0in
\epsfbox{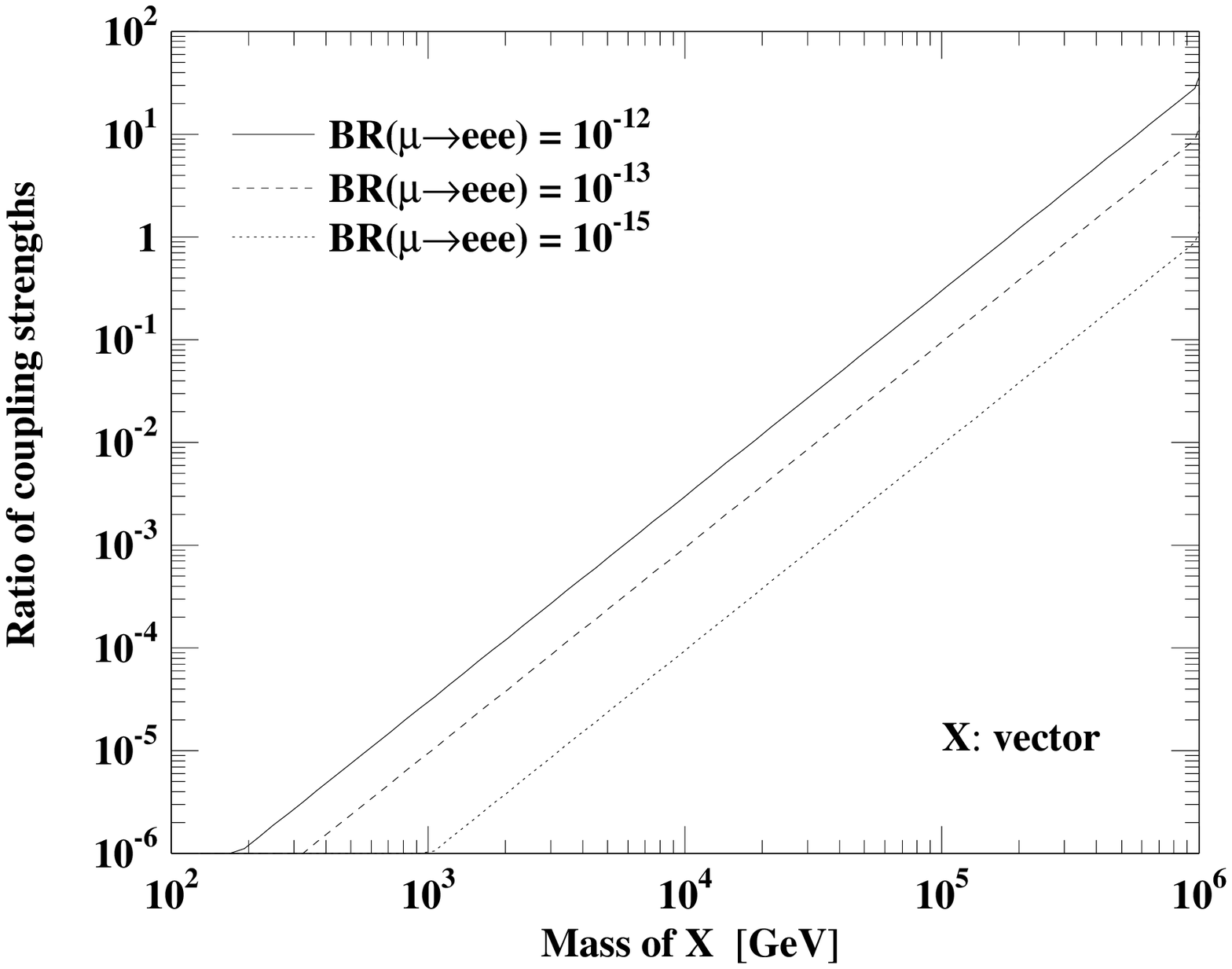}
\vspace{1.0cm}
\caption{The ratio of coupling strengths $g_{e\mu}^{^V}/g_{ee}^{^Z}$ 
  as a function of $M_X$ for various values of ${\cal BR}(\muToeee)$.
  In this plot, $X$ is assumed to be a vector.  
  For a scalar $X$, the shape
  is very similar but the coupling strength is about twice as big.
  Since the experimental upper limit for the branching ratio is
  $10^{-12}$, 
  the region above the solid curve is experimentally excluded.}
\label{fig1}
\end{figure}

In order to estimate the experimental sensitivity of 
$e \mu $ collision to the FCNC phenomena, we make an assumption 
for the coupling strengths using  the upper limit of experimental
branching ratio for the $\muToeee$ decay as a guide, 
${\cal BR}(\muToeee) < 1.0\times 10^{-12}$.  
For simplicity, we consider the case where $X$ is a vector.
{}From Eq.~(\ref{Gamvme3}), and the total decay width of muon
which is, in a good approximation, 
$\Gamma_{\rm total} (\mu) \simeq 
      \Gamma (\mu \rightarrow e \nu \bar{\nu}) =
{G_F^2 m_{\mu}^5 / 192 \pi^3}$, we get the branching ratio,
\begin{equation}
{\cal BR}(\muToeee) \simeq 16 
      \left( {\sqrt{g_{e\mu}^{^V} g_{ee}^{^V}}
\over {g_{ee}^{^Z}}} \right)^4 \left( M_Z \over M_X \right)^4. 
\end{equation} 
In addition, for simplicity, we assume the flavor conserving
coupling $g_{ee}^{^V}$ to be equal to $g_{ee}^{^Z}$, 
the electroweak electron coupling to $Z$ boson.  
Then the experimental upper limit gives a constraint 
for the coupling strength $g_{e\mu}^{^V}$ and the mass $M_X$;
\begin{equation}
{g_{e\mu}^{^V} \over g_{ee}^{^Z}} < {\sqrt{\cal BR} \over 4} 
\left( {M_X \over M_Z } \right)^2 .
\label{eq:BR} 
\end{equation}
Fig.~\ref{fig1} shows the ratio of coupling strengths
$g_{e\mu}^{^V} / g_{ee}^{^Z}$ as a function of $M_X$ for various
values of ${\cal BR}(\muToeee)$; 
$10^{-12}$ (solid curve), $10^{-13}$
(dashed curve), and $10^{-15}$ (dotted curve).  Since the present
experimental upper limit is $1.0\times 10^{-12}$ \cite{Bellgardt}, 
the region above the solid curve is experimentally excluded.

Note that the measurement of the decay width $\Gamma(\muToeee)$ alone, 
even if it were measured precisely, cannot determine the coupling 
strength $g^{^V}_{e\mu}$ and $M_X$ separately.
We clearly need a high-energy collision experiment so that we can 
scan the energy ranges and directly determine the mass and width of 
the new intermediate boson $X$. Only with the information on
the mass and width, we can determine the coupling strengths.

\begin{figure}
\leavevmode
\epsfxsize=6.25in
\epsfbox{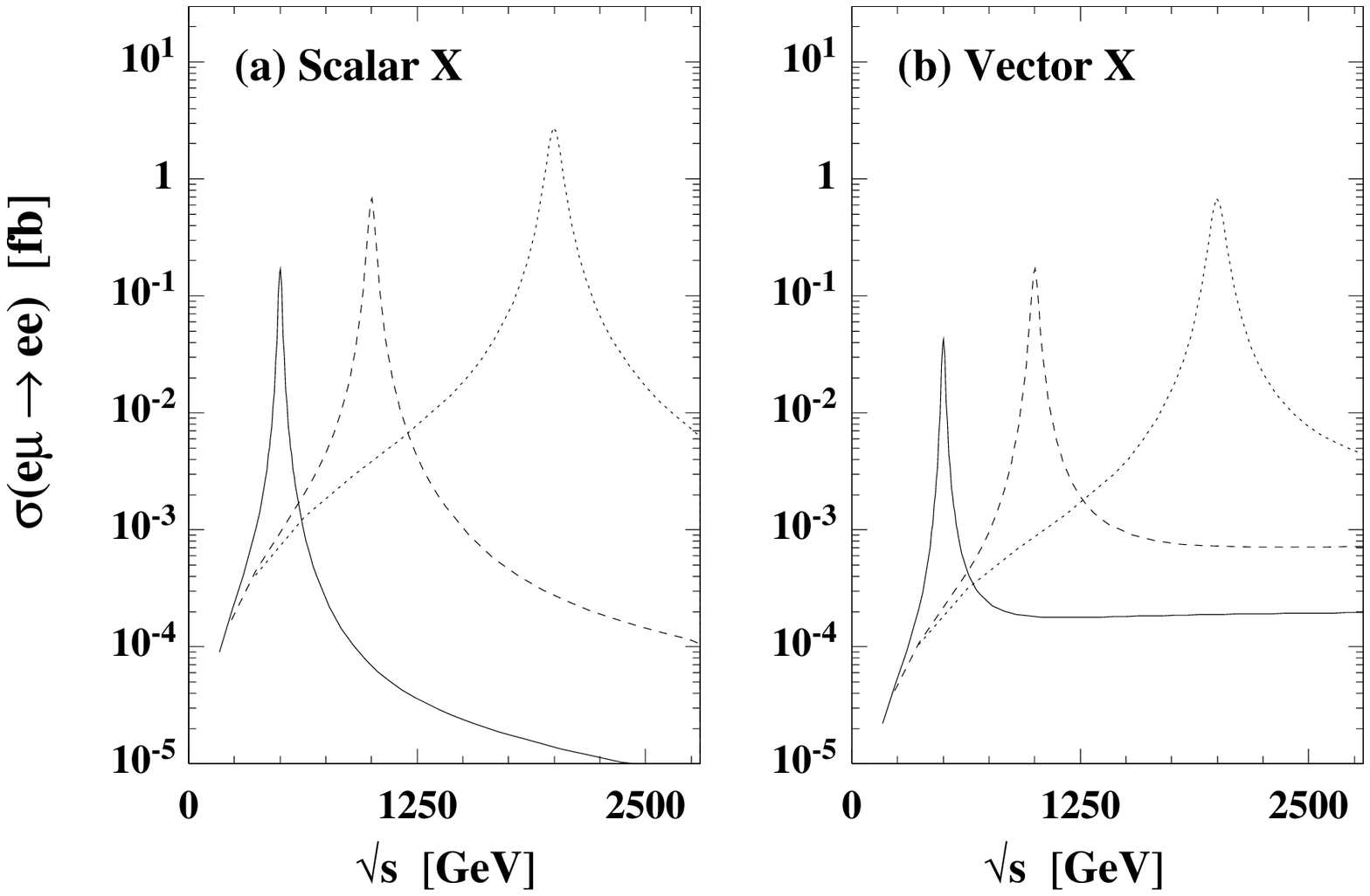}
\vspace{1.0cm}
\caption{The cross-section $\sigma (\mueToee)$ vs. $\sqrt{s}$ for 
  $M_X = 500,\ 1000,\ 2000~\GeV$: 
  (a) for $X$ being a scalar, and (b) for a vector, respectively.
  The $X$ width $\Gamma_X$ is assumed to be proportional to $M_X$, 
  and $\Gamma_X=20$ GeV is taken for $M_X=500$ GeV.}
\label{fig2}
\end{figure}

To make a simple quantitative estimate of the required 
luminosity for an $e \mu $ collider, we first choose the value of $M_X$, 
and then from Eq.~(\ref{eq:BR}) and Fig.~\ref{fig1} 
we decide coupling strengths by assuming a branching ratio to 
be an order of magnitude smaller than
the current experimental upper limit, 
{\it eg.} for $M_X = 500~\GeV $
we get $ g_{e\mu}^{^S} / g_{ee}^{^Z} = 5.50\times 10^{-6}$ and $
g_{e\mu}^{^V} / g_{ee}^{^Z} = 2.38\times 10^{-6}$, which corresponds
to ${\cal BR}(\muToeee) = 10^{-13}$ ({\it i.e.}  $1 \over 10$ of the
current experimental upper limit). 
The width of $X$ is chosen to be proportional\cite{Zmass}
to the mass $M_X$,
after taking $\Gamma_X=20~\GeV$ for $M_X=500$ GeV.
Fig.~\ref{fig2} shows the
calculated cross-section as a function of $\sqrt{s}$ for 
$M_X = 500,\ 1000,\ 2000~\GeV$, 
with the coupling strengths determined from
Fig.~\ref{fig1} assuming ${\cal BR}(\muToeee) = 10^{-13}$.  

With those preset assumptions of coupling strengths, mass and width
for a new neutral boson $X$, the evaluated peak cross-section is
found to be in the range of $1~{\rm fb}$ for $M_X \sim 1~\TeV$.  
As mentioned earlier, we emphasize again that
the FCNC signals can be depicted in the $e \mu $
collision via enormously many final-state channels.
Therefore, even if we conservatively count this large number of 
channels as a factor 
of 10 increase in the summed cross-section of visible FCNC channels, 
we expect to observe a significant number of FCNC events on the 
resonance peak with $1~{\rm fb^{-1}}$ integrated luminosity. 
Without a substantial background, that would be sufficient
to claim an experimental evidence for FCNC, but the background issue
shall be more carefully studied with details of detector and collider
design parameters.

On the other hand, Fig.~\ref{fig2} shows that the cross-section off
the resonance peak is typically about $\sim 10^{-2}~{\rm fb}$ for
$\sqrt{s} \sim 1~\TeV$.  Once again, if we count the multi-channel
final state as a factor 10 enhancement in the total visible
cross-section, this implies that an integrated luminosity of $100~{\rm
fb^{-1}}$ will be sufficient to experimentally observe quite a few
FCNC signals under low-background.  Note that the B-factory
experiments being prepared at SLAC \cite{SLAC-B} and KEK \cite{KEK-B}
are aiming at a luminosity of $100~{\rm fb^{-1}}/{\rm year}$.
Therefore, we conclude that if we can maintain the $e\mu$ collision
luminosity at the level of the B-factory experiments, and if the
intermediate boson $X$ has the properties that we have assumed, then
we may have a very good chance to observe FCNC signals in a few years
running.  The chance can be much enhanced if we run the experiment at
the right energy, {\it i.e.}  at or near $\sqrt{s} = {M_X}$.

In conclusion, we have investigated the physics potential of $e\mu$
collision by using a simple model calculation and demonstrated that
the $e \mu $ collision experiment can offer the best laboratory to
search for FCNC at high energies.  

\medskip
\medskip

\noindent
{\bf Note added:} After we finished the first version of this
manuscript, we found the similar work by Barger et al. \cite{barger},
which suggested to use a relatively low energy muon beam that may be
available during the first stages of muon collider to probe physics.
 
\begin{center}
{\bf Acknowledgments}
\end{center}
We would like to thank M. Drees, J. Hewett and P. Zerwas for helpful 
discussions, and P. Ko for bringing our attention to a talk by G.W.S. Hou.
SYC thanks the DESY theory group for its hospitality during his stay.
The work  was supported 
in part by Non-Directed-Research-Fund, KRF in 1997, 
in part by KOSEF through CTP of SNU, 
in part by Yonsei University Faculty Research Fund of 1997, 
in part by the BSRI Program, Ministry of Education, 
Project No. BSRI-97-2425,
and in part by the KOSEF-DFG large collaboration project, 
Project No. 96-0702-01-01-2.

%\pagebreak

\pagebreak


\begin{thebibliography}{99}

\bibitem{cronin}
  J.H. Christensen {\it et al.}, Phys. Rev. {\bf 126} (1962) 1202.

\bibitem{Huet}
  A.D. Sakharov, ZhETF Pis. Red. {\bf 5} (1967) 774; JETP Lett. {\bf 5}
  (1967) 24; G.R. Farrar and M.E. Shaposhnikov, Phys. Rev. D {\bf 50}
  (1994) 382; P. Huet and E. Sather, Phys. Rev. D {\bf 51} (1995) 359;
  A.G. Cohen, A. De R\'{u}jula and S.L. Glashow, 
  astro-ph/9707087.

\bibitem{g2hdm}
  P.~Sikivie, Phys. Lett. {\bf B65} (1976) 141;
  A.B.~Lahanas and C.E.~Vayonakis, Phys. Rev. {\bf D19} (1979) 2158;
  G.C.~Branco, A.J.~Buras and J.-M.~G\'erard,
  Nucl. Phys. {\bf B259} (1985) 306;
  J.~Liu and L.~Wolfenstein, Nucl. Phys. {\bf B289} (1987) 1;
  T.P.~Cheng and M.~Sher, Phys. Rev. {\bf D35} (1987) 3484;
  M. Sher and Y. Yuan, Phys. Rev. {\bf D44} (1991) 1461;
  A.~Antaramian, L.J.~Hall and A.~Ra\v sin, Phys. Rev. Lett. 
  {\bf 69} (1992) 1971;
  G. Cveti\v c, S.S. Hwang and C.S. Kim, hep-ph/9706323. 

\bibitem{col}
  M. Luke and M.J. Savage, Phys. Lett. {\bf B307} (1993) 387;
  W.S.~Hou, Phys. Lett. {\bf B296} (1992) 179;
  D.~Atwood, L.~Reina and A.~Soni, Phys. Rev. Lett. {\bf 75} (1995) 3800;
  N. Arkani-Hamed, H.C. Cheng, J.L. Feng and L.J. Hall,
  Phys. Rev. Lett. {\bf 77} (1996) 1937;
  S.~Bar-Shalom, G.~Eilam, A.~Soni and J.~Wudka, hep-ph/9703221.

\bibitem{Bolton} 
  R.D. Bolton {\it et al.}, Phys. Rev. D {\bf 38} (1988) 2077; 
  W. Honecker {\it et al.}, Phys. Rev. Lett. {\bf 76} (1996) 200.

\bibitem{Bellgardt} 
  U. Bellgardt {\it et al.}, Nucl. Phys. {\bf B299} (1988) 1;
  V.A. Baranov {\it et al.}, Sov. J. Nucl. Phys. {\bf 53} (1991) 802.

\bibitem{mam} 
  B.E. Matthias $et~al.$, Phys. Rev. Lett. {\bf 66} (1991) 2716.

\bibitem{hou}
  G.W.S. Hou, talk given at 3rd International Conference on Physics
  Potential and Development of $\mu^+ \mu^-$ Colliders, 
  San Francisco, CA, 13-16 Dec 1995. 
 
\bibitem{Barger} V. Barger, M.S. Berger, J.F. Gunion and T. Han,
  Phys. Rev. Lett. {\bf 75} (1995) 1462;``$\mu^+ \mu^-$ Collider: A
  Feasibility Study'', BNL-52503 (1996).

\bibitem{Zmass} 
The measured $Z$ boson total width is $2.49$ GeV and the width 
is proportional to the $Z$ boson mass. 
Because $X$ also decays mainly 
into two particles, it is reasonable to assume that the $X$ width
is proportional to its mass.

\bibitem{SLAC-B} 
  D. Boutigny {\it et al.} (BaBar Collaboration),  
  Technical Design Report, 1995.

\bibitem{KEK-B}
  M.T. Cheng {\it et al.} (Belle Collaboration), 
  Technical Design Report, KEK-Preprint 95-1, 1995; 
  KEK-Preprint 97-1, 1997.

\bibitem{barger} V. Barger, S. Pakvasa and X. Tata, hep-ph/9709265.

\end{thebibliography}
\end{document}